\def\breakon{\end{multicols}\widetext\vspace{-.6cm}
\noindent\rule{.49\linewidth}{.3mm}\rule{.3mm}{.5cm}\vspace{0.0cm}}
\def\breakoff{\vspace{-0.45cm}
\noindent
\rule{.50\linewidth}{.0mm}\rule[-.47cm]{.3mm}{.5cm}\rule{.49\linewidth}{.3mm}
\vspace{-0.25cm}
\begin{multicols}{2}   }
\documentstyle[aps,epsf,prb,multicol]{revtex}
\begin{document}

\makeatletter
\renewenvironment{table}
  {\let\@capwidth\linewidth\def\@captype{table}}
  {}

\renewenvironment{figure}
  {\let\@capwidth\linewidth\def\@captype{figure}}
  {}
\makeatother

\title{Conductivity of Doped Two-Leg Ladders}
\author{Eugene H. Kim}
\address{Department of Physics, University of California,  
          Santa Barbara, California 93106-9530  \\ and  \\
         $^\dag$ Department of Physics and Astronomy, 
         McMaster University, Hamilton, Ontario, Canada L8S-4M1}          
\maketitle

\begin{abstract}
Recently, conductivity measurements were performed on the
hole-doped two-leg ladder material ${\rm Sr_{14-x}Ca_xCu_{24}O_{41}}$.
In this work, we calculate the conductivity for doped two-leg
ladders using a model of hole-pairs forming a strongly
correlated liquid -- a single component Luttinger liquid -- in
the presence of disorder.  Quantum interference effects are 
handled using renormalization group methods.  We find that our 
model can account for the low energy features of the experimental 
results.  However, at higher energies the experiments show 
deviations from the predictions of this model.  Using the results 
of our calculations as well as results on the ground state of doped 
two-leg ladders, we suggest a scenario to account for the higher 
energy features of the experimental results.  
\end{abstract}


\vspace{.15in}
\begin{multicols}{2}

Over the last few years, there has been considerable interest in 
ladder materials.\cite{rice}  In particular, the two-leg ladder 
has received quite a bit of attention.  Theoretically, the two-leg 
ladder is popular becase one can do controlled calculations, both
analytically and numerically.  Experimentally, two-leg ladders are 
popular because there are several known two-leg ladder materials,
and these materials can be fabricated in a controlled way.  In 
particular, the two-leg ladder material 
 ${\rm Sr_{14-x}Ca_xCu_{24}O_{41}}$ has received a considerable 
amount of attention.\cite{mccarron}  Unlike some of the other ladder 
materials which must be fabricated under high pressure, this material 
can be fabricated under ambient pressure.  Also, when ${\rm Ca}$ replaces 
some of the ${\rm Sr}$, holes are doped into the ladders.  However, 
these holes were already present in the material in the charge
reservoir layers composed of ${\rm CuO_2}$ chains; doping with 
 ${\rm Ca}$ simply moves them to the ladders.  Therefore, this 
material is self-doped, and hence, extremely clean.  Besides these 
practical reasons, the two-leg ladder is interesting because its 
properties are remarkably similar to the cuprate superconductors, 
namely when doped with holes, the system has a spin gap and the 
doped holes form pairs with a $d_{x^2-y^2}$-like internal structure.   

Recently, conductivity measurements were performed on the 
hole-doped two-leg ladder material ${\rm Sr_{14-x}Ca_xCu_{24}O_{41}}$. 
 \cite{osafune}  The $c$-axis resistivity, $\rho_c$, 
(i.e. the resistivity parallel to the chains of the ladder)
is insulating at low temperatures and shows metallic $T$ dependence 
above $T_0 \approx 100K$ for $x=11$; $\rho_c$ is nonmetallic over 
the entire region $T < 300K$ for $x=8$.  The $a$-axis resistivity, 
$\rho_a$, (i.e. the resistivity parallel to the rungs of the ladder) 
is insulating at low temperatures and shows a gradual change of 
sign in the $T$ coefficient at 
 $T^* \approx 250K$ for $x=11$; $\rho_a$ is insulating for $x=8$.  
(See Fig.~1 of Ref.~\ref{osafune}.) 

The $a$-axis optical conductivity, $\sigma_{1a}(\omega)$, shows 
pseudogap behavior in the low frequency region, the depressed 
spectral weight being transferred into the higher energy region.  
(See Fig.~2 of Ref.~\ref{osafune}.)  
The $c$-axis optical conductivity, $\sigma_{1c}(\omega)$, shows 
a low frequency peak.  Unlike a Drude peak, the peak in 
$\sigma_{1c}(\omega)$ is located at a finite frequency, $\omega_0$: 
$\omega_0 \sim 100cm^{-1}$ for $x=8$ and $\omega_0 \sim 50cm^{-1}$ 
for $x=11$.  (See Fig.~3 of Ref.~\ref{osafune}.)

In this work, we calculate the conductivity for doped two-leg
ladders using a model of hole-pairs forming a strongly
correlated liquid.  We focus on the $c$-axis conductivity, but we 
make some remarks about the $a$-axis conductivity.  Our model is 
motivated by several known facts.  First, the two-leg ladder 
(at half-filling and for not too large dopings) is known to have a 
spin gap.  Second, the holes doped into the two-leg ladder are known 
to form pairs with an internal  $d_{x^2-y^2}$-like structure.  
Therefore, for our (effective low-energy) model, we consider a liquid 
of $d_{x^2-y^2}$ hole pairs in the presence of (weak) disorder.  Due 
to the one-dimensional nature of the ladder, the pairs do not condense.  
Rather, they form a strongly correlated liquid --- a single-component 
Luttinger liquid.  
                                                             
It should be noted that our model is not purely phenomenological.  
At weak coupling, controlled calculations using bosonization have 
been done.\cite{lin,orignac}  In these calculations, one gets 
four bosonic modes --- two spin modes and two charge modes.  These 
calculations show that, for not too large dopings, both of the spin
modes are gapped and one of the charge modes is gapped.  Therefore, 
the low-energy physics is described by a single gapless bosonic 
charge mode.  Numerical work\cite{noack} shows that there is no 
phase transition as one goes from weak to strong coupling.  Therefore, 
the low-energy effective theory in terms of a single gapless bosonic 
charge mode is still valid.  However, the parameters of this effective 
theory cannot be computed as they can in weak coupling; they must be 
taken as phenomenological parameters from numerics.  

The Hamiltonian we consider is
\begin{eqnarray}
 H & = & \frac{v}{2} \int dx~ 
     \left[K \pi^2 + \frac{1}{K} (\partial_x \phi)^2\right]  
    \nonumber \\
   & + & \int dx \left[ \xi(x) e^{-i\sqrt{4\pi}\phi}
       + \xi^*(x) e^{i\sqrt{4\pi}\phi} \right]  \, .
\label{themodel}
\end{eqnarray}
This Hamiltonian was first considered in Ref.~\ref{orignac}, 
and in that work they computed the resistivity, $\rho_c(T)$.  In 
this work, we extend their calculation of $\rho_c(T)$ to lower 
temperatures, and we also calculate the optical conductivity, 
$\sigma_{1c}(\omega)$.  In Eq.~\ref{themodel}, $\phi$ and $\pi$ 
are conjugate fields satisfying $[\phi(x),\pi(y)]=i\delta(x-y)$.
 $\xi(x)$ and $\xi^*(x)$ are Gaussian random variables\cite{abrikosov} 
satisfying
\begin{equation}
 \overline{\xi^*(x_1) \xi(x_2)} 
          = \Delta_{\cal D}~ \delta(x_1 - x_2) \, ,
\end{equation}
where the overbar denotes disorder averaging.  
$K$, the Luttinger liquid parameter, is determined
by the interactions.  For attractive interactions, $K > 1$; for 
repulsive interactions, $K < 1$.  ($K = 1$ for a non-interacting system.)
Note that the coupling to disorder is via the $4k_F$ charge density 
wave (CDW).\cite{orignac}  Due to the gaps in the spin modes and one 
of the charge modes, the $2k_F$ CDW coupling is irrelevant; the most 
relevant coupling generated is to the $4k_F$ CDW. 
For later convenience, we define a dimensionless disorder strength,
${\cal D}$, \cite{kane}
\begin{equation}
 {\cal D} = \frac{1}{n_i E_0^2}~ \Delta_{\cal D} \, ,
\end{equation}
where $n_i$ is the density of impurities and $E_0$ is a high-energy
cut-off.  Since our model is a liquid of $d_{x^2-y^2}$ hole pairs, 
$E_0 \sim$ pair binding energy.


We begin with a standard renormalization group (RG) analysis.
In an RG treatment we coarse-grain our system, integrating out
high energy degrees of freedom, and deriving an effective theory
for the low energy degrees of freedom.  The RG equations for the
parameters can be obtained in a straightforward way using standard
techniques.\cite{cardy}  They are
 \begin{eqnarray}
 \frac{d {\cal D} }{dl} & = & (3-2K)D  \, , \nonumber \\
 \frac{d K }{dl} & = &  - D K^2  \, ,   \\
 \frac{d v }{dl} & = &  - D K v  \, .  \nonumber 
\end{eqnarray}
A few words are in order about the RG equations.   
From the RG equation for ${\cal D}$, we see that there is a 
localization-delocalization transition at $K=3/2$.  However, for the 
systems we are considering, $K < 1$.  Hence, ${\cal D}$ is 
strongly relevant and flows to strong coupling; all states will be 
localized at low temperatures.  Also note that $K$ as well as the 
speed of the excitations, $v$, decrease under the RG.  


From the Kubo formula, the $c$-axis conductivity for our model is 
given by 
\begin{equation}
 \sigma_{1c}(\omega,T) = \frac{1}{\omega}~
    {\rm Im} \langle \overline{J(x,t) J(x',t')} \rangle(q=0,\omega) 
    \, ,  
\label{kubo}
\end{equation}
where $J(x,t)$ is the current operator 
\[
 J(x,t) = \frac{v}{\sqrt{\pi}}~ \partial_t \phi \, .
\]
Fourier transforming in space and time, we have
\begin{equation}
  \sigma_{1c}(\omega,T) = - \frac{v^2 \omega}{\pi} 
    {\rm Im} \langle \overline{\phi(x,t) \phi(x',t')}
    \rangle(q=0,\omega)   \, . 
\end{equation}

From the RG analysis, we saw that the disorder grows strong at
low energies and the system localizes.  In this work, we will
not be concerned with the properties of the localized (insulating) 
state.  More formally, we will be interested in the properties near 
the Luttinger liquid (free boson) fixed point.  Therefore, we can 
use RG techniques\cite{zinnjustin,amit} to compute the 
conductivity.  Consider the Green's function
\begin{equation}
   G(\omega, {\cal D}) = \langle \overline{\phi(x,t) \phi(x',t')} 
                         \rangle(q=0,\omega)   \, .
\end{equation}
Suppose we scale energies --- $\omega \rightarrow b\omega$ ($b=e^{dl}$).  
The Green's function transforms as
\begin{equation}
 G(\omega, {\cal D}) = b^2 Z~ G(b\omega, b^{3-2K} {\cal D}) 
 \, .
\end{equation}
The factor $b^2$ comes from dimensional analysis, while the factor
 $Z$ is due to the ``wave function renormalization'' which arises
in the RG.  For simplicity, we ignore the renormalization of
 $K$ and $v$.  Since, by assumption, the disorder is weak 
and since we are far from the localization-delocalization transition, 
the renormalization of $K$ and $v$ should be small.\cite{orignac}
Writing $Z = b^{-\gamma}$, where $\gamma = {\cal D}K$, and iterating 
gives
\begin{equation}
  G(\omega, {\cal D}) = 
  \exp\left[\int_0^{l^*} dl \left(2 - \gamma(l)\right) \right]
  G\left( e^{l^*}\omega, e^{(3-2K)l^*}{\cal D} \right)  \, .
\label{RGimproved}
\end{equation}
We choose $l^*$ such that $e^{l^*} = (E_0/E)$, where $E = \omega$
or $T$.

Notice that on the right hand side of Eq.~\ref{RGimproved}, $G$ 
is evaluated at $E_0$.  At such high energies, quantum interference 
effects should be negligible.    
In this regime, the Born approximation gives accurate results for 
the conductivity.  In the Born approximation, the Green's function 
has the form
\begin{equation}
 G(\omega,{\cal D}) = \frac{1}{\omega^2 + i\omega/\tau}  \, .
\label{borngreens}
\end{equation}
The functional form of $\tau$ can be deduced from scaling
arguments.\cite{giaschulz,kane}  In the Born approximation, 
 $1/\tau \sim n_i v~ {\cal D}$.  Also, from scaling
\begin{equation}
  \frac{1}{\tau(b E, b^{3-2K}{\cal D})} = 
  \frac{b}{\tau( E, {\cal D})}  \, ,
\end{equation}
where $E$ = $\omega$ or $T$.  Therefore, we deduce that
\begin{equation}
  \frac{1}{\tau(E, {\cal D})} = 
  {\rm const} \times n_i v~ {\cal D} \left( \frac{E_0}{E} \right)^{2-2K}  \, .
\label{lifetime}
\end{equation}
Knowing the conductivity in the Born approximation, we can use 
Eq.~\ref{RGimproved} to relate the conductivity at lower energies, 
where quantum interference effects are important, to the conductivity 
at higher energies, where quantum interference effects are negligible
and the Born approximation is accurate.


Let us first understand what our expressions tell us about the 
resistivity, $\rho_c(T)$.  
Using Eqs.~\ref{borngreens} and \ref{lifetime} with 
Eq.~\ref{RGimproved} gives 
\begin{equation}
 \rho_c(T) \sim \left(\frac{1}{x}\right)^{2-2K}
           \exp \left\{ \frac{K{\cal D}}{3-2K} 
           \left[ \left(1/x\right)^{3-2K} - 1 \right] \right\}
\label{resistivity}
\end{equation}
where we have introduced $x = T/E_0$.
Consider the first factor on the right hand side.  This power-law
factor can be deduced by scaling arguments.\cite{giaschulz}
The second factor arises from the ``wave function renormalization''
in the RG and is due to quantum interference/localization 
effects.  


Now let us consider the optical conductivity, $\sigma_{1c}(\omega)$
(at $T=0$).  Again, using Eqs.~\ref{borngreens} and \ref{lifetime} 
with Eq.~\ref{RGimproved} gives 
\begin{equation}
 \sigma_{1c}(\omega) \sim \frac{x^{2-2K}}{x^{6-4K} + \delta^2}
  \exp\left\{ \frac{-K{\cal D}}{3-2K} 
          \left[ \left(1/x \right)^{3-2K} - 1 \right] \right\}
\label{conductivity}
\end{equation} 
where $\delta \sim n_i v~{\cal D}/E_0$ and we have introduced 
$x = \omega/E_0$.  Consider the first factor on the right hand side.  
Notice that for $K=1$, we have a peak at $\omega=0$ --- a Drude peak.  
For $K < 1$ (i.e. for an interacting system with repulsive 
interactions), the peak is shifted to finite frequency.  The 
smaller the value of $K$, the higher the frequency at which the
peak occurs.  Now consider the second factor.  This factor arises from
the ``wave function renormalization'' in the RG.
This factor is due to quantum interference/localization effects, and 
it suppresses the low frequency behavior.


\vspace{.2in}
\begin{figure}
\epsfxsize=3.0in
\centerline{\epsfbox{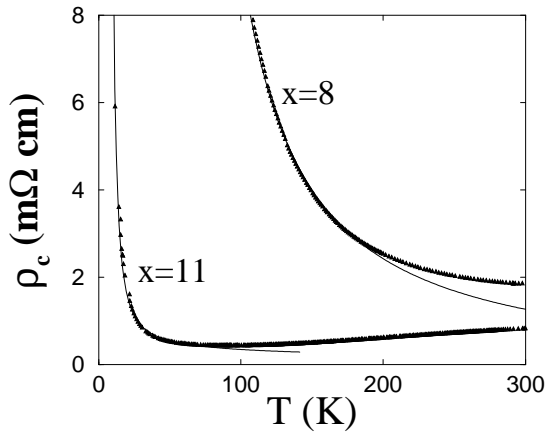} }
\vspace{.2in}
\caption{ $\rho(T)$ vs. $T$. Solid lines are from 
 Eq.~\ref{resistivity}.  The experimental data is from
 Ref.~\ref{osafune}. }
\label{fig:vstheoryrhoplot}
\end{figure}
\vspace{.2in}

How do our results compare with the experimental results of
Ref.~\ref{osafune}?  In Fig.~\ref{fig:vstheoryrhoplot}, we
plot our calculated result for $\rho_c(T)$ as well as the 
experimental results of Ref.~\ref{osafune}.  To fit the data 
for $x=8$, we took $K=.25$, ${\cal D} = 0.1$, $E_0 = 425K$, 
and $\rho_0 = 0.74 m\Omega~cm$.  To fit the data for $x=11$, we
took $K=0.78$, ${\cal D} = 0.1$, $E_0 = 142K$, and
$\rho_0 = 0.285 m\Omega~ cm$.  
For $x=8$, we were able to get a reasonable fit to the data at 
lower temperatures.  However, we cannot account for the rather
constant resistivity starting at $\sim 200K$.  We believe that
pairing fluctuations and the renormalization of $K$ and $v$, which 
we ignored in our calculation, could account for this difference.  
For $x=11$, we were able to get a reasonable fit to the data at low 
temperatures.  However, our calculated result cannot account for 
the metallic behavior seen above $T_0 \approx 100K$.  We believe
the metallic behavior is due to pairing fluctuations, which were
not included in our model.  (See below for more details on pairing
fluctuations.)

\vspace{.2in}
\begin{figure}
\epsfxsize=3.2in
\centerline{\epsfbox{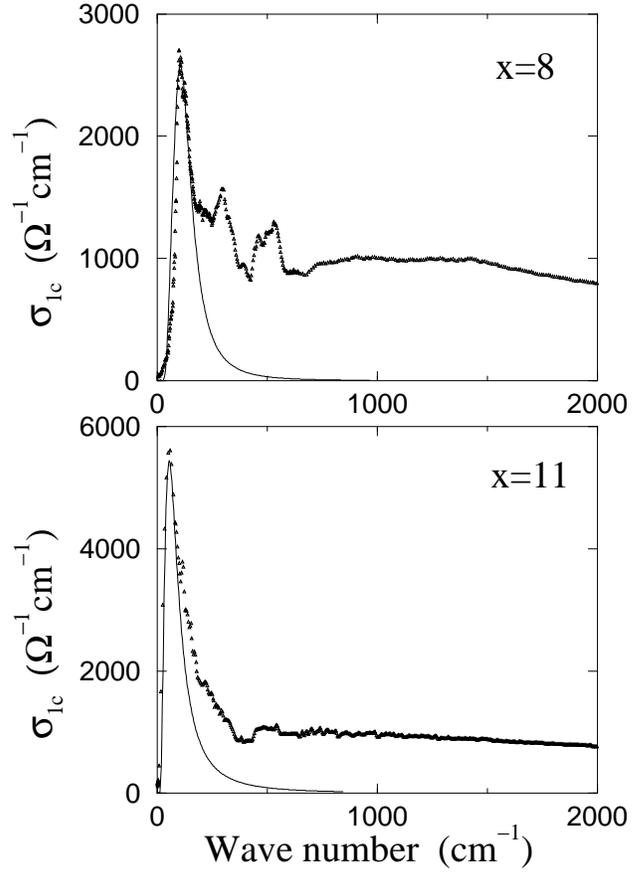} }
\vspace{.2in}
\caption{$\sigma_1(\omega)$ vs. Wave number.  Solid lines are
 from Eq.~\ref{conductivity}.  The experimental data is from 
 Ref.~\ref{osafune} at $10K$. }
\label{fig:sigmaplot}
\end{figure}
\vspace{.2in}

In Fig.~\ref{fig:sigmaplot} we plot our calculated result for 
$\sigma_{1c}(\omega)$ as well as the experimental results of
Ref.~\ref{osafune}.  We see that $\sigma_{1c}(\omega)$ shows a
narrow, finite frequency peak.  This finite frequency (rather
than $\omega=0$) peak is due to the strong correlations; the 
smaller the value of $K$, the larger the frequency at which the 
peak occurs.  However, at higher energies, our calculated result 
does not reproduce some of the features of the measured conductivity.  
Namely, there is appreciable spectral weight at higher energies, 
while the calculated result gives practically no spectral weight 
at higher energies.  We believe this appreciable spectral weight 
at higher energies is due to pairing fluctuations, which were not 
included in our model.  
In order to fit the data, we took $K=0.25$, ${\cal D}=0.1$, 
$\delta=0.075$, $E_0 = 0.04cm^{-1}$, and 
 $\sigma_0 = 150\Omega^{-1}cm^{-1}$ for x=8; we took $K=0.78$, 
${\cal D}=0.1$, $\delta=0.1$, $E_0 = 0.035cm^{-1}$, and 
 $\sigma_0 = 355\Omega^{-1}cm^{-1}$ for x=11.
The values for $E_0$ and $\sigma_0$ are inconsistent with the 
values determined from the resistivity.
For $x=8$, the peak occurs at $\sim 100cm^{-1}$ $(=144K)$.
At these energies the renormalization of $K$ and $v$, which we
ignored in our calculation, could be coming into play.  For 
$x=11$, the peak occurs at $\sim 50cm^{-1}$ $(=72K)$.  At these 
energies, pairing fluctuations are probably still be present and 
could account for our inability to fit the data consistently.  
(See below for more details on pairing fluctuations.)

\begin{figure}
\epsfxsize=3.25in
\centerline{\epsfbox{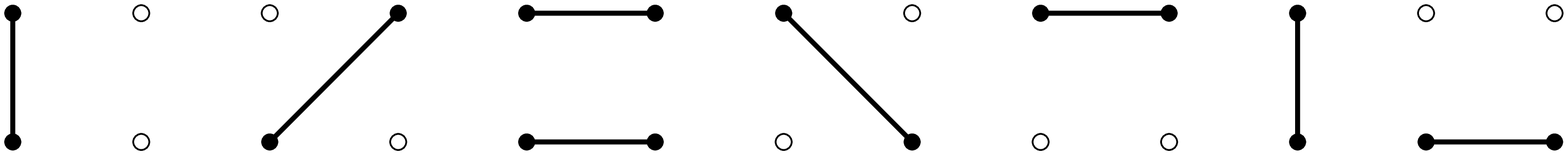} }
\vspace{.2in}
\caption{ Typical configuration in the ground state of doped two-leg
ladders -- hole pairs moving through an RVB background.  (Figure
taken from Ref.~\ref{sierra}.) }
\label{fig:RVB}
\end{figure}

For not too large dopings, a simple picture which captures the
essential physics of the ground state of doped two-leg ladders
is hole pairs moving through a nearest-neighbor resonating 
valence bond (RVB) background.\cite{sierra}  A typical configuration 
in the ground state is shown in Fig.~\ref{fig:RVB}.  In this picture, 
the spin gap is related to the energy necessary to break the short 
ranged singlet bonds; pairing also occurs in order to minimize the 
number of broken singlet bonds.  From this picture of the ground
state, combined with our calculations for the conductivity, we 
suggest a scenario to explain the experimental results of 
Ref.~\ref{osafune}.  

\vspace{.2in}
\vspace{.2in}

$T_0$ $(\approx 100K)$ is the temperature at which the $c$-axis 
resistivity becomes metallic for $x=11$.\cite{osafune}  We believe 
that this is the temperature at which pairing fluctuations occur.  
In other words, the RVB background is still, for the most part, 
in tact.  However, the holes are no longer tightly bound into 
pairs.  In the bosonization treatment, pairing is due to 
the ``pinning'' of $\theta_{\rho^-} = \theta_{\rho,b} - \theta_{\rho,a}$,
the relative phase of the charge modes in the bonding and antibonding 
bands.\cite{lin,orignac}  When $\theta_{\rho^-}$ is pinned, the relative 
phase between the bonding and antibonding bands become locked together,
with fluctuations being gapped.  In our calculation, we took 
 $\theta_{\rho^-}$ to be pinned; we ignored fluctuations, since 
they are gapped.
We speculate that above $T_0$, $\theta_{\rho^-}$ is no longer pinned
and is free to fluctuate.  
This also explains why our calculated conductivity has almost no
spectral weight at higher energies.  Namely, at higher energies 
the missing spectral weight is due to the fluctuations of 
$\theta_{\rho^-}$ which we ignored.  

$T^*$ $(\approx 250K)$ is the temperature at which the $a$-axis 
resistivity becomes metallic for $x=11$; it was found to coincide 
with the spin-gap inferred from NMR measurements.\cite{osafune}
From the ground state wave function, we see that it is energetically 
costly for single electrons to tunnel between neighboring ladders as
long as the RVB background is in tact i.e. as long as there is a 
spin-gap.    
Above $T^*$, the spin-gap is destroyed, and hence it is not costly 
for single electrons to tunnel between neighboring ladders.  In the 
bosonization language, due to the spin gap, single particle tunneling 
is irrelevant.  However, the operator which tunnels a pair between 
neighboring ladders is generated under the RG.  It is the most relevant 
operator generated, and hence dominates the $a$-axis transport.  Once 
the spin-gap is destroyed, the $a$-axis transport will be dominated
by single particle tunneling. 

$\sigma_{1a}(\omega)$ was observed to exhibit pseudogap behavior.
The pseudogap was identified to open up at $\sim 1100cm^{-1}$ 
$(=1584K)$ for $x=8$ and $ \sim 600cm^{-1}$ $(=864K)$ for 
 $x=11$.\cite{osafune}  These values are much larger than the 
spin-gap values.  However, it is possible that the energies at 
which the pseudogap begins were overestimated in 
Ref.~\ref{osafune}.\cite{timusk}  We believe that the pseudogap 
should coincide with the spin-gap, namely because that is where 
single particles become free to tunnel between neighboring ladders.


To summarize, we computed the conductivity of doped two-leg ladders.  
Our model consisted of hole pairs forming a single component Luttinger 
liquid in the presence of disorder;  
quantum interference effects were handled using RG methods.
Our results could account for  
the low-energy features of Ref.~\ref{osafune}.  However, there were 
some discrepancies at higher energies, namely we could not account 
for the metallic behavior in $\rho_c(T)$ for $x=11$, and we could not 
account for the significant spectral weight in $\sigma_{1c}(\omega)$ 
at higher energies.  Based on the results of our calculations as well 
as results on the ground state of doped two-leg ladders, we suggested 
that these discrepancies were due to pairing fluctuations which were 
not included in our model.  We also commented that it is quite natural 
that $T^*$ corresponds to the spin-gap inferred from NMR measurements; 
single particles are free to tunnel between ladders above $T^*$, but 
not below $T^*$. 
Finally, we commented that we believe the
energy at which the pseudogap begins in $\sigma_{1a}(\omega)$ should
coincide with the spin-gap, and that its value may have been 
overestimated in Ref.~\ref{osafune}.


The author is especially grateful to T. Osafune, N. Motoyama,
H. Eisaki, S. Uchida, and S. Tajima for use of their data; to 
N. Motoyama for correspondence; and to T. Timusk for innumerable
helpful discussions about the data.
The author would also like to thank A.~J. Berlinsky, C. Kallin, 
D.~J. Scalapino, and especially K.~M. Kojima and J. S\'{o}lyom 
for helpful discussions and comments about the manuscript.  
Finally, the author gratefully acknowledges the hospitality 
of Argonne National Laboratory, where parts of this manuscript 
were written.  This work was supported by the DOE under grants
No. 85-ER45197 and No. W-31-109-ENG-38.



\end{multicols}
\end{document}